\documentclass[12pt]{elsarticle}
\usepackage{amssymb,amsmath,amsthm}
\usepackage{rotating}
\usepackage{graphicx,comment}
\usepackage{color}

\newtheorem{proposition}{Proposition}

\theoremstyle{definition}
\newtheorem{remark}{Remark}

\excludecomment{omitext}

\begin{document}

\title{ 
An ill-posed problem in hydrodynamic  stability of multi-layer 
Hele-Shaw flow
}

\author{Gelu Pa\c{s}a}  
\address{Simion Stoilow Institute of Mathematics of Romanian  Academy \\
Calea Grivi\c tei 21, Bucharest 010702, Romania} 

\ead{gelu.pasa@imar.ro}

\begin{abstract}
An useful approximation for the displacement of two immiscible  fluids in 
a porous medium  is the Hele-Shaw  model.  
We consider several  liquids  with different constant viscosities, inserted 
between  the displacing fluids.
The linear stability analysis of this model leads us to an ill-posed  
problem. The growth  rates (in time) of the perturbations exist iff  some 
compatibility conditions on the interfaces are verified. We prove that
these conditions   cannot be fulfilled.
         \end{abstract}


\maketitle

\section{Introduction}

The  Stokes flow in a rectangular Hele-Shaw cell is studied in  \cite{BE}, \cite{HS},  
\cite{LAMB}. In \cite{SAFF-TAY} has been shown that the sharp interface, which exists 
between two immiscible  fluids in a Hele-Shaw  cell  (or an equivalent  porous medium), 
becomes  unstable if the displacing fluid  is less viscous.

Some experimental and numerical results (see \cite{PASA}  and references therein) have 
shown that
an intermediate liquid  with  a continuous {\it variable} increasing viscosity in the 
flow direction,  inserted  between the  displacing fluids, can minimize  the  Saffman -
Taylor instability. 
A theoretical model for this  three-layer flow  was first considered in \cite{GOR-HOM-1}, 
\cite{GOR-HOM-2}.  An optimal intermediate viscosity, which minimizes the instability, 
was obtained  by using  a numerical procedure. 

In a large number of papers was considered the multi-layer Hele-Shaw flow: the injection 
of a sequence of fluids,  with different viscosities, in a homogeneous equivalent porous 
medium. Some instability control seems   be obtained by  using this flow model - see   
\cite{PRAB-TIPM},    \cite{PRAB-SIAM}   and references therein.

 We consider the following problem:  can we use a sequence of liquids with {\it constant} 
viscosities, inserted between the initial fluids,  in order to minimize the Saffman-Taylor 
instability ?

In this paper we give a negative answer for the above poblem.
The analysis of the linear  stability  of this  model leads us to an ill-posed problem. The growth 
rates (in time) of the perturbations  are obtained by using usual  boundary  conditions on the 
interfaces,  based on the Laplace-Young law. We show that these conditions, related also to the 
amplitude of the velocity perturbations  in each layer, cannot be fulfilled. Thus the growth 
rates can not  exist.

\section{ The flow model }

In  \cite{GOR-HOM-1}, \cite{GOR-HOM-2}   was proposed an ``optimal policy''  in order  to 
minimize the Saffman-Taylor instability,  which  appears   when a less viscous fluid is 
displacing a more viscous one  in a  Hele-Shaw cell.  
The point is   to  consider an {\it intermediate} region,
 filled by  a given amount  of polymer-solute, with a variable continuous viscosity.
This ``three-layer''  structure is moving with the velocity, say,  $U$, of the displacing fluid
far upstream. 
The cell is parallel with  the plane ${\overline{x}}Oy$ and the flow is in the 
positive direction $O{\overline{x}}$.
A basic solution is considered, with two straight interfaces. The length of intermediate 
region (between interfaces) is constant.
The viscosity $\mu({\overline{x}})$ of the intermediate liquid is invertible with respect to the 
concentration of the polymer-solute. The continuity equation for the polymer-solute is used to get
 the equation of the (a priori) unknown viscosity $\mu({\overline{x}})$  in the intermediate region:
\begin{equation}\label{VISC}
 \mu_t + u\mu_{\overline{x}} + v\mu_y =0,
\end{equation}
where $(u,v)$ are the velocity components and the indices $t,{\overline{x}}, y$ denote  the partial 
derivatives with respect to time and ${\overline{x}},y$. 
In \cite{GOR-HOM-1}, \cite{GOR-HOM-2} were  considered a viscosity jump and a surface tension {\it only 
on the second interface}, between  the intermediate region  and the displaced fluid. On the first 
interface, the viscosity was continuous and the surface tension was missing. 
A linear stability analysis of the basic slolution was performed and  an ``optimal'' viscosity  was 
obtained by using a numerical procedure. This optimal viscosity gives us  the minimum values 
of  the growth rates (in time) of perturbations.

If  the intermediate region is large enough, the viscosity is continuous  and surface tensions are 
 missing  on both interfaces, then there  exist a variable  viscosity which leads us to an 
arbitrary small (positive) growth rates of perturbations  - see \cite{PASA}.

In this paper   we consider a  basic solution  with several  constant intermediate  viscosities. The  
surface  tensions exist  on all  interfaces. As in  \cite{GOR-HOM-1}, \cite{GOR-HOM-2}, the Laplace-Young 
law is used to get the boundary   conditions on the  interfaces (which depends also on the amplitudes of 
the perturbations of the horizonthal velocity). We show that these conditions cannot be fulfilled. Thus 
the growth rates can not   exist.

We recall below the model given in \cite{GOR-HOM-1}, \cite{GOR-HOM-2}, with      variable 
intermediate   viscosity  and use  it to get the flow model with   constant intermediate  
viscosity.

The  following three-layer basic flow with the intermediate region  ${\overline{x}} \in 
(Ut+a,Ut+b)$   is considered 
$$
 u=U, \,\, v=0; \quad   {\overline{x}}_L = Ut+a, \quad {\overline{x}}_R =Ut+b;            $$
$$ \quad P_{x}= -\mu_G U, \quad \quad  P_y=0;                                             $$
$$
\mu_G= \mu_L, \,\, {\overline{x}} <{\overline{x}}_L,  \quad   \quad 
\mu_G = \mu_R, \,\, {\overline{x}} >{\overline{x}}_R,                                     $$
\begin{equation}\label{ZT004A}
 \mu_G=\mu({\overline{x}}-Ut), \,\, {\overline{x}} \in 
({\overline{x}}_L, {\overline{x}}_R), \quad                                                                       
\end{equation}
where $P$ is the basic pressure,  $\mu_G$ is the basic  viscosity,
$\mu_L, \mu_R$ are the constant viscosities of the displacing and displaced fluids. The last 
relation  is obtained by using the relations  \eqref{VISC} with  $u=U, v=0$.
We use   the moving reference frame $x=  {\overline x} - Ut$.  Thus  the intermediate region 
becomes the segment  $(a,b)$, in which  $\mu_G= \mu(x)$.  
{\it  Without loss of generality we can consider  $b < 0$. }

The stability system for small perturbations, obtained in \cite{GOR-HOM-1}, is  linear. Thus 
the following    perturbation   $u'$ of the horizontal velocity is considered
\begin{equation}\label{FOURIER-U}
 u'(x,y,t) =                                                       
f(x)  [\cos(ky) +i \sin (ky)] e^{\sigma t}, \,\, k \geq 0,  \quad i=\sqrt-1, 
\end{equation}
where   $f(x)$ is the amplitude, $\sigma $ is the growth constant  and  $k$ 
are the wave numbers.
The cross derivation of the perturbed pressures  leads us to  the amplitude 
equation (see the relation (2.17) of \cite{GOR-HOM-1}, where $f$ is denoted 
by $\psi$   and $\mu$ is denoted by $\mu_0$)
\begin{equation}\label{ZT008}
 -(\mu f_x)_x +  k^2 \mu f = \frac{1}{\sigma} U k^2 f \mu_x,  
 \quad \forall x \notin \{a,b \}.
\end{equation}
The viscosity is constant outside the intermediate region, thus it follows
$$
 -f_{xx} + k^2 f = 0, \quad x \notin (a,b);                                    $$ 
\begin{equation}\label{amplitude}
  -(\mu f_x)_x +  k^2 \mu f = \frac{1}{\sigma} U k^2 f \mu_x,  
 \quad \forall x \in (a,b).            
\end{equation}
The perturbations must decay to zero in the far field and $f$ is continuous, 
therefore we get
\begin{equation}\label{FAR-FIELD}
f(x) = f(a) e^{  k(x-a) }, \,\, \forall x \leq a; \quad 
f(x) = f(b) e^{ -k(x-b) }, \,\, \forall x \geq b.
\end{equation}

\vspace{0.25cm}

As we mentioned above, he boundary conditions are obtained by using the Laplace - 
Young  law. It means the 
pressure jump (across the interfaces) should equal  the surface tension $T$ 
times  the curvature  and the horizontal velocity should be  continuous.

Let's consider   two surface tensions $T(a), T(b)$ and two viscosity jumps  
at $x=a$ and $x=b$. Thus  $f_x$ could be discontinuous  in $a,b$. The 
right and left limit values at   the points $a,b$ are  denoted by the 
superscripts $^+, ^-$. Just like in  \cite{GOR-HOM-1}, page 84 formula 
(2.27),  we get:
$$  \mu^-(a) f_x^-(a) - \mu^+(a) f_x^+(a)= \frac{kE(a)}{\sigma}f(a),  $$
$$  \mu^-(b) f_x^-(b) - \mu^+(b) f_x^+(b)= \frac{kE(b)}{\sigma}f(b),  $$ 
$$ 
E(a):= k U [\mu^+(a)-\mu_L] - k^3T(a),                                   $$ 
\begin{equation}\label{LAPLACE002}
E(b):= k U [\mu_R-\mu^-(b)] - k^3T(b).   
\end{equation}
In both relations \eqref{LAPLACE002}  we have the same eigenvalue(s), 
then  the  {\it unknown} amplitude $f$ must verify the compatibility 
relation
$$  
\frac{kE(a)f(a)} {\mu^-(a) f_x^-(a) - \mu^+(a) f_x^+(a)}  =            $$
\begin{equation}\label{LAPLACE003} 
\frac{kE(b)f(b)} {\mu^-(b) f_x^-(b) - \mu^+(b) f_x^+(b)}.
\end{equation} 
The stability of the basic solution  \eqref{ZT004A} is governed   by 
the  system  \eqref{amplitude} - \eqref{LAPLACE003}.

\section{The non existence of the growth rates}

We consider a constant intermediate     viscosity $\mu= \mu_1$ s.t. 
$$
0<  \mu_L < \mu_1 < \mu_R,                                           $$
\begin{equation}\label{HYP1}
\mu^-(a)=\mu_L, \quad  \mu^+(a)=\mu_1=\mu^-(b), \quad \mu^+(b)=\mu_R.
\end{equation}  
Thus both  viscosity jumps are positive in the flow direction. As  $\mu$ is 
constant inside the intermediate region, from \eqref{amplitude} we obtain
\begin{equation}\label{CONST} 
 - f_{xx} +  k^2  f = 0,  \quad \forall x \in (a,b).                                            
\end{equation}
The stability system for the flow with constant viscosity \eqref{HYP1} is 
given by the equations  \eqref{FAR-FIELD}-\eqref{CONST}.

A crucial difference from the Gorell and Homsy model exists:  the  solution of 
\eqref{CONST} is {\it independent} of $\sigma$. This property leads us to  the 
non-existence of the eigenvalues. We recall \eqref{FAR-FIELD} - \eqref{HYP1}  
and  we  introduce the notations
\begin{equation}\label{LAPLACE002A} 
VL:=  \frac{kE(a)f(a)}{ \mu_L kf (a) - \mu_1 f_x^+(a)}, 
\quad  
VR:=  \frac{kE(b)f(b)}{ \mu_1 f_x^-(b) +k \mu_R f (b)},                            
\end{equation}
Therefore the compatibility condition \eqref{LAPLACE003} is equivalent with $VL=VR$. 

\vspace{0.25cm}

\begin{proposition}
 The growth constants   of the problem   \eqref{FAR-FIELD}-\eqref{CONST} can not exist.
To this end, we prove that the condition \eqref{LAPLACE003}  cannot be  fulfilled.

{\it Proof}. Suppose that there exist  growth   rates of the considered problem.
 The solution of \eqref{CONST} is $f(x)=A e^{kx}+Be^{-kx}$, where $A,B$ can depend on 
$k$ and $|f(x)| < \infty,  \,\,\,\forall k$.  Thus the  relations  \eqref{LAPLACE002}   
become 
$$ \mu_L( Ae^{ka}+Be^{-ka})- \mu_1 ( Ae^{ka}- Be^{-ka}) = 
\frac{E_a}{\sigma}( Ae^{ka}+Be^{-ka}),                                               $$
$$ \mu_1( Ae^{kb}-Be^{-kb})+ \mu_R ( Ae^{kb}+ Be^{-kb}) = 
\frac{E_b}{\sigma}( Ae^{kb}+Be^{-kb}),                                               $$
and we  get the   system 
$$ A e^{ka}(\mu_L-\mu_1-\frac{E_a}{\sigma})+
Be^{-ka}(\mu_L+\mu_1-\frac{E_a}{\sigma})=0,                                           $$
\begin{equation}\label{F-2}  
   A e^{kb}(\mu_1+\mu_R-\frac{E_b}{\sigma})-
Be^{-kb}(\mu_1-\mu_R-\frac{E_b}{\sigma})=0.    
\end{equation}
A solution  $(A,B)\neq (0,0)$ exists if the following   condition is verified: 
$$
 e^{k(a-b)}(\mu_L-\mu_1-\frac{E_a}{\sigma})(\mu_ 1-\mu_R-\frac{E_b}{\sigma})+        $$                   
\begin{equation}\label{F-3}
 e^{k(b-a)}(\mu_1+\mu_R-\frac{E_b}{\sigma})(\mu_L+\mu_1-\frac{E_a}{\sigma})=0.
\end{equation}
We study  the values of the growth rates  when $k$ is large enough, such that 
$e^{2 k(a-b)} \approx 0$ (recall $a-b <0, k > 0  $).  Thus
\begin{equation}\label{F-4}  
(\mu_1+\mu_R-\frac{E_b}{\sigma})(\mu_L+\mu_1-\frac{E_a}{\sigma})=0.
\end{equation}
The  relation \eqref{F-4} is  a second order equation for $\sigma$ and we have two real 
roots:
\begin{equation}\label{F-5} 
{\sigma_1}= \frac{ E_b}{\mu_1+\mu_R}, \quad {\sigma_2}= \frac{ E_a}{\mu_L+\mu_1}.
\end{equation}

We insert $\sigma_1$   in   $\eqref{F-2} _2$ and get $2 \mu_R B e^{-kb} =0$.  We have   
 $\mu_R >0$, then it follows $B=0$. So actually the solution to the equation \eqref{CONST} is  
$f(x)=  Ae^{kx}$. 
We use the notations \eqref{LAPLACE002A}, we impose  the condition $VL=VR$, then we get  
\begin{equation}\label{F-7}
\frac{k U (\mu_1-\mu_L) - k^3T(a)}{\mu_L-\mu_1}  = \frac{k U (\mu_R-\mu_1) - k^3T(b)}{\mu_1+\mu_R} .                                          
\end{equation}
We equate the coefficients of $k, k^3$ and get
\begin{equation}\label{F-8}
\mu_R =0, \quad  \mu_1[T(a)+T(b)]= \mu_L T(b).                                    
\end{equation}    
From the relation \eqref{F-7} with  large $k$   we obtain 
\begin{equation}\label{F-9}
 \frac{T(a)}{T(b)} = \frac{\mu_L-\mu_1}{\mu_1+\mu_R} \Rightarrow   \mu_L >\mu_1.
\end{equation}
The relationship $\eqref{F-8}_1$ and $\eqref{F-9}_2$  contradict the hypothesis \eqref{HYP1}.

We  insert $\sigma_2$  in $\eqref{F-2}_1$ and get $2 \mu_1 A e^{ka} =0$. 
The viscosity $\mu_1$  must be strictly positive, thus $A=0$ and the solution of \eqref{CONST} 
becomes  $f(x)= B e^{-kx}, \,\, x<0, k >0 $. 
The values of $f(x)$ must be finite, thus we impose the condition $ \max_k Be^{-kx}<  \infty $.  
We use the notations \eqref{LAPLACE002A}   with $f(x)=Be^{-kx}$, we impose  the condition 
$VL=VR$, then it  follows  
\begin{equation}\label{F-7A}
\frac{k U (\mu_1-\mu_L) - k^3T(a)}{\mu_L+\mu_1}  = \frac{k U (\mu_R-\mu_1) - k^3T(b)}
{-\mu_1+\mu_R} .                                          
\end{equation}
We equate the coefficients of $k, k^3$ and get
\begin{equation}\label{F-8A}
\mu_L =0, \quad  \mu_1[T(a)+T(b)]= \mu_R T(a).                                    
\end{equation}    
The condition \eqref{F-7A} for  large enough $k$ gives us
\begin{equation}\label{F-9A}
 \frac{T(a)}{T(b)} = \frac{\mu_L+\mu_1}{\mu_R-\mu_1},
\end{equation}
which is a restriction on the viscosities. The  relations \eqref{F-8A}, \eqref{F-9A}
contradict the hypothesis \eqref{HYP1}.
The bottom line is: the condition \eqref{LAPLACE003} is not satisfied, thus the problem 
\eqref{FAR-FIELD} - \eqref{CONST} has no eigenvalues. 
\end{proposition}    \hfill $\square$

%%%%%%%%%%%%%%%%%%
\begin{remark}
We consider now a constant intermediate  viscosity $\mu= \mu_1$ s.t. 
$$  0<  \mu_L > \mu_1, \quad \mu_1  < \mu_R.                                        $$

We get the same relationship \eqref{F-8} - \eqref{F-9} and \eqref{F-8A} -  \eqref{F-9A}.
This time, the relationship $\mu_L > \mu_1$  is not a contradiction, but $\mu_R=0$ and
$\mu_L=0$ contradicts the hypothesis \eqref{HYP1}.   
For large $k$, the relations   \eqref{F-9} and \eqref{F-9A}    represent restrictions on 
the viscosity $\mu_1$, which also contradict  the assumption \eqref{HYP1}. 
\end{remark}   \hfill $\square$

%%%%%%%%%%%%%%%%%%

\vspace{0.25cm}

We   use the above results for the case of   $N$ intermediate layers.
Consider $(N+1)$ interfaces $x_i$ and $N$  constant viscosities $\mu_i$ 
in  each layer $(x_{i-1}, x_i)$:
$$   x_0 = a< x_1 < x_2  < ...< x_N =b,                              $$
\begin{equation}\label{VISCO}
0< \mu_L \equiv \mu_0 < \mu_1< \mu_2 <...<\mu_i ... < \mu_N  <  \mu_R.         
\end{equation}
The surface tensions, amplitudes, viscosities and limit values in the 
points  $x_i$ are denoted  by 
$$
T_i=T(x_i); \quad f_i=f(x_i);                                         $$ 
$$
f_x^{+,-}(i)=f_x^{+,-}(x_i); \quad 
\mu^{+,-}(i)=\mu^{+,-}(x_i);                                          $$
\begin{equation}\label{VISCO-3}
 \mu(x)= \mu_i, \,\, \forall x \in (x_{i-1}, x_i); \quad  
 \mu_i = \mu^+(i-1)= \mu^-(i).                                            
\end{equation}
The  stability of the flow  model with viscosities \eqref{VISCO}  is 
governed by the relations \eqref{FAR-FIELD} and the system 
\eqref{N-0}-\eqref{N-0A1}  
below:
\begin{equation}\label{N-0} 
-f_{xx}+k^2f=0, \quad x \in (a,b), \quad x \neq x_i;
\end{equation}
$$
\mu^-(i)  f_x^+(i) - \mu^+(i) f_x^+(i) =  \frac{kE_if_i}{\sigma},    $$
\begin{equation}\label{N-01}
E_i = {k U [\mu^+(i)-\mu^-(i)] - k^3T_i};
\end{equation}
\begin{equation}\label{N-0A}
 V_0=V_1=...= V_i = ...= V_N;
\end{equation}
\begin{equation}\label{N-0A1}
V_i \equiv \frac{kE_if_i}{ \mu^-(i)  f_x^+(i) - \mu^+(i) f_x^+(i)}.
\end{equation}
 
\vspace{0.25cm}

\begin{proposition} 
The above conditions \eqref{N-0A} cannot be fulfilled.                                  

{\it Proof.} 
i) The solutions of the equations  \eqref{N-0} in the intervals  $(x_{i-1}, x_1)$ could 
be  
$$ f_i(x)=A_i e^{kx}+B_ie^{-kx},                                                     $$ 
where $ A_i,B_i $ are  absolute constants. 
However, for large $k$, the terms $|B_ie^{-kx}|$  become very large. Thus  $|u'|$
also becomes very large and we exceed the frame of small perturbations. The linear 
stability analysis makes no sense. For this reason, we consider {\it only} the solutions
$ f_i(x)=A_i e^{kx}$,  where  $A_i$ are  absolute   constants.    The amplitude  $f$ is 
continuous at the points $x_i$. That means 
$$ A_i \exp(kx_i) = A_{i+1}\exp(kx_i), \quad \forall k   \Rightarrow  A_i=A_{i+1},  \quad 
\forall i .                                                                            $$
Therefore $A_i$  are the same for all $ x\in [a,b]$  and exists a  constant, say,  $A$ 
such that $A_i=A,\quad \forall i $.  

 In the following, we show that the compatibility relation  $V_0=V_N$ can not 
be fulfilled. We will highlight some  restrictions on the viscosities $\mu_i$ which  contradict  
the  hypothesis   \eqref{VISCO}.

We impose the condition $V_0=V_N$ with  $f(x)=A  e^{kx}$, thus  
 \begin{equation}\label{N-6}
\frac{k U (\mu_1-\mu_L) - k^3T_0}{\mu_L-\mu_1}  = \frac{k U (\mu_R-\mu_N) - k^3T_N}{\mu_N+\mu_R} .                                          
\end{equation}  
We equate the coefficients of $k, k^3$ and get $ \mu_R =0, \quad \mu_NT_0+\mu_1 T_N= \mu_L T_N$,                                    
in contradiction   with \eqref{VISCO}. 
The equation  \eqref{N-6} with large enough $k$ gives us 
$T_0/T_N = (\mu_L-\mu_1)/(\mu_N+\mu_R)$, so  $\mu_L>\mu_1$, which also  contradict the hypothesis 
\eqref{VISCO}.

\vspace{0.25cm}

ii)  
The most general  solution of \eqref{N-0} is  
$$
f(x) = A_i e^{kx}+Be_i^{-kx}, \quad x \in  (x_{i-1}, x_i),                        $$ 
\begin{equation}\label{N-1A} 
A_i=A_i(k),  \,\, B_i=B_i(k),\quad  1 \leq i \leq N.
\end{equation}  
In order to remain in the frame of small perturbations, the maximal  
values and the  limits of  $A_ie^{kx}, B_ie^{-kx}$ (for large  $k$) 
must be finite. It seems natural to impose the following conditions:
$$
\max_k \{ A_ie^{kx}  \} < \infty,  \,\,
\lim_{k \rightarrow \infty} A_ie^{ kx}=   AI< \infty;                            $$
\begin{equation}\label{k-large}
\max_k \{ B_ie^{-kx} \} < \infty, \,\, 
\lim_{k \rightarrow \infty} B_ie^{-kx}= BI< \infty.
\end{equation}
We have to prove $V_0 \neq V_N$. To this end, we introduce the notations 
$$A=A_1, \quad B=B_1, \quad C=A_N, \quad D=B_N,                                  $$
   $$ f(x)= A(k) e^{kx}+B(k)e^{-kx}, x \in (x_0, x_1);                           $$
   $$ f(x)= C(k) e^{kx}+D(k)e^{-kx}, x\in (x_{N-1}, x_N).                        $$
Suppose  $V_0=V_N$, then 
\begin{equation}\label{N-1B}
 \frac{E_0}{E_N} =  \frac{\mu_L-\mu_1 X} {\mu_NY+\mu_R},                       
\end{equation}
$$ X=\frac{Ae^{ka}-Be^{-ka}}{Ae^{ka}+Be^{-ka}}, \quad 
   Y = \frac{Ce^{kb}-De^{-kb}}{Ce^{kb}+De^{-kb}},                                $$
$$F_1:= \lim_{k \rightarrow \infty}X = \frac{A_1-B_1}{A_1+B_1}, \quad 
  F_N:= \lim_{k \rightarrow \infty}Y = \frac{A_N-B_N}{A_N+B_N}.                   $$
We consider  large values of $k$ in \eqref{N-1B}, the above  relations gives us  
$$ \frac{T_0}{T_N} =  \frac{\mu_L-\mu_1F_1} {\mu_NF_N+\mu_R}.                     $$
This is a restriction on the viscosities $\mu_i$ which   contradicts  the hypothesis  
\eqref{VISCO}. If $B_1=B_N=0$, then  we recover the formula \eqref{F-9}.

Our conclusion is following:  the  conditions \eqref{N-0A} cannot be fulfilled,  thus  
the eigenvalues of the problem  \eqref{N-0}-\eqref{N-0A1}  cannot exist.   
We use   {\it Remark 1} and get the same result for negative viscosity jumps in the
flow direction                               
\end{proposition}   \hfill $\square$

\begin{remark}
In the last part of \cite{PASA} was studied the linear stability of the flow with an 
intermediate liquid  with {\it continuous  viscosity and  without  surface tensions}; 
we obtained  
$\mu^+(a)=\mu_L,  \,\,  \mu^-(b)=\mu_R, \,\, T(a)=T(b)=0.                          $
The growth rates appear only in the equation \eqref{amplitude} and {\it not}  in the 
boundary conditions \eqref{LAPLACE002}. Both terms in \eqref{LAPLACE003} become zero.
\end{remark}    \hfill  $\square$

\end{document}